\newif\iffigs\figstrue
\documentclass[11pt]{article}
\usepackage{latexsym,amssymb}
\iffigs
   \input{epsf}
\else
\fi
\newcommand{\ft}[2]{{\textstyle\frac{#1}{#2}}}

\def\1bar{1\hskip -.275cm -}
\def\2bar{2\hskip -.275cm -}
\def\3bar{3\hskip -.275cm -}

\newsavebox{\uuunit}
\sbox{\uuunit}
    {\setlength{\unitlength}{0.825em}
     \begin{picture}(0.6,0.7)
        \thinlines
        \put(0,0){\line(1,0){0.5}}
        \put(0.15,0){\line(0,1){0.7}}
        \put(0.35,0){\line(0,1){0.8}}
       \multiput(0.3,0.8)(-0.04,-0.02){10}{\rule{0.5pt}{0.5pt}}
     \end {picture}}

\makeatletter \@addtoreset{equation}{section} \makeatother

\def\bfone{\relax{\rm 1\kern-.35em 1}}

\def\bfone{\relax{\rm 1\kern-.35em 1}}
 
\begin{document}
\begin{titlepage}
\begin{center}
{\LARGE \bf   BPS D3-branes on smooth}\\
{\LARGE \bf  ALE manifolds$^\dagger$}\\ \vfill
{Talk given at the Conference {\it New Trends in Particle Physics} \\
September 2001, Yalta, Ukraina}
\vfill
{\large
 Pietro Fr\'e} \\
\vfill { Dipartimento di Fisica Teorica, Universit\'a di Torino, $\&$ INFN -
Sezione di Torino\\
via P. Giuria 1, I-10125 Torino, Italy  }
\end{center}
\vfill
\begin{abstract}
{In this talk I review the recent construction of a new family of classical BPS solutions of type IIB
supergravity describing 3-branes transverse to a 6-dimensional space
with topology  $\mathbb{R}^{2}\times$ALE. They are characterized by a
non-trivial flux of the supergravity 2-forms through the homology $2$-cycles of
a generic smooth ALE manifold.
These solutions have two Killing spinors and thus preserve $\mathcal{N}=2$
supersymmetry. They
are expressed in terms of a quasi harmonic function $H$
(the ``warp factor''), whose properties was studied in detail in the case
of the simplest ALE,
namely the Eguchi-Hanson manifold. The equation for $H$
was identified as an instance of the confluent Heun equation.

}
\end{abstract}
\vspace{2mm} \vfill \hrule width 3.cm {\footnotesize $^ \dagger $
This work is supported in part by the European Union RTN contracts
HPRN-CT-2000-00122 and HPRN-CT-2000-00131.}
\end{titlepage}

\section{Introduction}
\label{intro}
After the seminal paper by Maldacena \cite{mal},
many efforts have been devoted to extend the gauge/gravity
correspondence to less supersymmetric and non-conformal cases. In
this context considerable attention was recently directed to
the study of fractional branes
\cite{Klebanov:2000rd}-\cite{marco}. These are the natural
elementary D-branes occurring whenever string theory is reduced on
a (not necessarily compact) orbifold \cite{Anselmi:1994sm},
\cite{Douglas:1996sw}-\cite{Diaconescu:1998br}.
Because of their nature they cannot
move away from the orbifold apex and thus the dual gauge
theory on their world-volume lacks the relevant moduli fields.
Generically, this leads to both reduced supersymmetry and
non-vanishing $\beta$-functions. Most interesting are the fractional
D3-branes, namely the case when the world-volume theory is
four-dimensional. In this respect the two most appealing
situations are provided by the ${\cal N}=1$ case emerging from
singular limits of CY spaces  and the $\mathcal{N}=2$ case arising
from the singular limit of ALE spaces. Much work was devoted to
both.
\par
A common feature of many supergravity solutions representing
non-conformal situations is the presence of naked singularities of
repulson type \cite{kal}. These correspond to IR singularities at
the gauge theory level and one expects that they should be resolved or
explained by some stringy effect. Although a general recipe
does not seem to exist, progress in understanding such an issue was
made both for the ${\cal N}=1$ and the $\mathcal{N}=2$ case.
\par
In the present talk I review recent work done in collaboration with
other authors \cite{paperus} where it was found that the bulk
solution of type IIB supergravity corresponding to fractional $D3$
can be generalized to the situation where the transverse space to the
brane has a smooth regular geometry and the topology of
$\mathbb{R}^2 \, \times \, ALE$ the last factor in this product
denoting a resolved Asymptotically Locally Euclidean $4$--manifold.
In our solution that is shown to be supersymmetric and hence describe
a BPS state of string theory there is a non
zero value of the complex type IIB supergravity 3-form in the transverse
directions. Note that this is the distinctive
feature of fractional D-branes in the singular orbifold theories.
Our bulk supergravity solution is a warped solution, but differently
from the case of usual branes the warp factor depends on two rather
than one radial variables and obeys a
  complicated partial differential equation. Actually the whole
set of IIB equations can be reduced to the solution of
such an equation for the warp factor, whose source is essentially
dictated by supersymmetry up to an arbitrary analytic
function $\gamma(z)$.
\par
Supergravity alone is not sufficient to determine $\gamma(z)$ or the
boundary conditions.
This arbitrariness implies that our solution describes various
deformation or various vacua of ${\cal N}=2$ theories. The case
$\gamma(z)= \hbox{const.}$ describes a vanishing three-form flux and
corresponds to the well known conformal
${\cal N}=2$ theory with product gauge group $\mathrm{U(N)} \times \mathrm{U(N)}$,
hyper-multiplets in the bi--fundamental representation and Fayet-Iliopoulos terms describing
the ALE moduli. The new ingredient in the construction of \cite{paperus} is the following. It was shown
how, at the supergravity level, a three-form flux
can be turned on consistently with supersymmetry.  Consequences of
this for the dual gauge theory are the goal of a new investigation
that is still ongoing.
\section{Bosonic action and field equations of type IIB supergravity}
\label{bosact2b}
As it is well known, type IIB supergravity does not have any conventional
supersymmetric action. However, as it happens for all on-shell
supergravity theories, the complete set of field equations can be
obtained as consistency conditions from the closure of the supersymmetry
transformation algebra \cite{Schwarz:1983qr}. In the case of type
IIB supergravity,  one was also able \cite{igorleo} to obtain a
complete, manifestly $\mathrm{SU(1,1)}$-covariant formulation of the theory
based on the rheonomic approach to supergravity theory
\cite{castdauriafre}.
\par
The bosonic part of the equations can be formally obtained through
variation of the following action \footnote{Note that our $R$ is
equal to $- \ft 1 2 R^{old}$, $R^{old}$ being the normalization of
the scalar curvature usually adopted in General Relativity
textbooks. The difference arises because in the traditional
literature the Riemann tensor is not defined as the components of
the curvature $2$-form $R^{ab}$ rather as $-2$ times such
components.}:
\[
S_{\rm IIB} = \frac{1}{2 \kappa^2} \Bigg\{ \int d^{10} x~ \left[
-2  \sqrt{-\det g}~ R \right] - \frac{1}{2} \int \Big[ d \varphi
\wedge \star d \varphi
 \,+\, {\rm e}^{- \varphi} F_{[3]}^{NS}  \wedge \star F_{[3]}^{NS}\,+\, {\rm e}^{2
 \varphi}\, F_{[1]}^{RR} \wedge \star F_{[1]}^{RR}
\]
\begin{equation}
 + \,\,{\rm e}^{\varphi} \,{F}_{[3]}^{RR} \wedge \star
 {F}_{[3]}^{RR} \,+\, \frac{1}{2}\, {F}_{[5]}^{RR}
 \wedge \star {F}_{[5]}^{RR}  \, -\,  C_{[4]} \wedge
 F_{[3]}^{NS}
 \wedge F_{[3]}^{RR} \Big] \Bigg\}
\label{bulkaction}
\end{equation}
where:
\begin{equation}
\begin{array}{cclcccl}
F^{RR}_{[1]} & = & dC_{[0]} &;& F^{NS}_{[3]} & = & dB_{[2]} \\
F^{RR}_{[3]}& = & dC_{[2]} -  \, C_{[0]} \,
dB_{[2]} & ; &
F^{RR}_{[5]}& = & dC_{[4]}- \ft 12 \left( B_{[2]} \wedge d C_{[2]}
-  C_{[2]} \wedge d B_{[2]}\right) \
\end{array}\label{bosecurve}
\end{equation}
It is important to stress though that the action
(\ref{bulkaction}) is to be considered only a book keeping device
since the $4$-form $C_{[4]}$ is not free, its field strength
$F_{[5]}^{RR}$  being subject to the on-shell self-duality
constraint:
\begin{equation}
F_{[5]}^{RR} = \star F_{[5]}^{RR} \label{selfonshell}
\end{equation}
>From the above action the corresponding equations of motion can be
obtained:
\begin{eqnarray}
d \star d \varphi - e^{2\varphi} \, F^{RR}_{[1]} \wedge \star
F^{RR}_{[1]} & = & -\ft 1 2 \, \left( e^{-\varphi} F^{NS}_{[3]}
\wedge \star  F^{NS}_{[3]}-
  e^{\varphi} F^{RR}_{[3]} \wedge \star  F^{RR}_{[3]}\right) \label{NSscalapr}\\
  d\left( e^{2\varphi} \star F^{RR}_{[1]}\right)  & = & - e^{\varphi} \, F^{NS}_{[3]} \wedge \star  F^{RR}_{[3]}
\label{RRscalapr}\\
d\left( e^{-\varphi} \, \star F_{[3]}^{NS}\right) + e^\varphi \,
F^{RR}_{[1]} \wedge \star F^{RR}_{[3]}
  & = &  - F_{[3]}^{RR} \wedge F^{RR}_{[5]}
\label{3formNS}\\
d\left( e^\varphi \star F_{[3]}^{RR } \right) & = & -F_{[5]}^{RR}
\, \wedge F_{[3]}^{NS}
\label{3formRR}\\
d\star F^{RR}_{[5]} & = & -
  F^{NS}_{[3]} \, \wedge \, F^{RR}_{[3]}
\label{f5RR}\\
-\,2 \,R_{{MN}}&=& \frac{1}{2}\partial_{{M}}\varphi
\partial_{{N}}\varphi+\frac{e^{2\varphi}}{2}
\partial_{{M}} C_{[0]} \partial_{{N}}
C_{[0]}+150
 {F}_{[5]{M}\cdot\cdot\cdot\cdot}
{F}_{[5]{N}}^{\phantom{{M}}\cdot\cdot\cdot\cdot}
\nonumber\\
& &+ 9 \left( e^{-\varphi}F_{[3]{M}\cdot\cdot}^{NS}\,
F_{[3]{N}}^{{NS}\phantom{{M}}\cdot\cdot} +e^{\varphi}{
F}_{[3]{M}\cdot\cdot}^{RR}
{ F}_{[3]{N}}^{RR\phantom{{M}}\cdot\cdot}\right)\nonumber\\
& & -\frac{3}{4}\,
g_{{MN}}\,\left(e^{-\varphi}F_{[3]\cdot\cdot\cdot}^{NS}
F_{[3]}^{NS\cdot\cdot\cdot}+e^{\varphi}{{F}}_{[3]\cdot\cdot\cdot}^{RR}{
F}^{RR\cdot\cdot\cdot}_{[3]}\right) \label{einsteinequa}
\end{eqnarray}
It is not difficult to show, upon suitable identification of the
massless superstring fields, that this is the correct set of
equations which can be consistently obtained from the manifestly
$\mathrm{SU(1,1)}$ covariant formulation of type IIB supergravity
\cite{igorleo}.

\section{$D3$-brane solution with ALE flux}
\label{3}
In this section we provide the BPS solution corresponding to a
3-brane transverse to a smooth ALE space, namely we construct type IIB
supergravity solutions describing 3 branes on a vacuum
$\mathbb{R}^{1,3} \times \mathbb{R}^{2}\times \mathrm{ALE}$. This will be
achieved without an analysis of the specific form of the
world-volume action of the brane, i.e. of the source term. Our
physical assumption will just be that, together with the usual RR
5-form flux, the 3-brane solution has a non-trivial flux of the
supergravity 2-form potentials along (one of) the compact two cycle(s)
of the blown-up orbifold (this translates into a non-vanishing value of the complex 3-form
field strength).

\subsection{Solution of the bosonic field equations}
\label{sol3flux}
We separate the ten coordinates of space-time into the following subsets:
\begin{equation}
 x^M = \left \{ \begin{array}{rcll}
&x^\mu \quad \mu =0,1,2,3& \mbox{coordinates of the 3-brane world volume}   \\
&z  \;=\;  x^4 + {\rm i} x^5 & \mbox{complex coordinate of
$\mathbb{R}^2 \sim \mathbb{C}$} \null \\
&y^\tau \quad \tau=6,7,8,9 & \mbox{real coordinates of the ALE
4-space }   \
\end{array} \right.
\label{coordisplit}
\end{equation}
and we make the following ansatz for the metric:
\begin{equation}
ds^2=H^{-\frac{1}{2}}\left (-\eta_{\mu\nu}dx^\mu\,dx^\nu \right
)+H^{\frac{1}{2}}dzd\bar{z}+H^{\frac{1}{2}}
ds^2_{ALE}\label{ansazzo}
\end{equation}
where the warp factor $
H=H(z,\bar{z},y)$ depends in principle on all the transverse
variables and  $ds^2_{ALE} = g^{ALE}_{\tau\sigma} dy^\tau\,dy^\sigma$
 is the metric of any ALE space and we
denote $ M_6 = \mathbb{R}^2 \times \mathrm{ALE}$ the six-manifold spanned by $z$, $\bar z$ and
$y^\tau$.
Defining
$
{\mathcal{H}}_\pm  =  \pm 2 \,e^{-\varphi/2}
F^{NS}_{[3]} + {\rm i} 2 \,e^{\varphi/2} \,F^{RR}_{[3]}
$,
 eq. (\ref{f5RR}) for the 5-form becomes:
\begin{equation}
d\star F^{RR}_{[5]} = {\rm i} \, \ft 1 {8} \,
{\mathcal{H}}_+ \wedge {\mathcal{H}}_-
\label{f5equazia}
\end{equation}
\par
Besides assuming the structure (\ref{ansazzo}) we also assume that
the two scalar fields, namely the dilaton $\varphi$ and the
Ramond-Ramond $0$-form $C_{[0]}$ are constant and vanishing $
\varphi=0 \, ; \, C_{[0]}=0$.
This assumption simplifies considerably the
equations of motion.
\par
The basic ansatz  characterizing our  solution and  providing  its
interpretation as a 3-brane with a flux  through a homology
2-cycle of the ALE space is given by the following:
\begin{equation}
\begin{array}{ccccccc}
  {\mathcal{H}}_+ & = & 2 \,d \gamma_I (z,{\bar z} )\, \wedge \,
  \omega^I & ; &
  {\mathcal{H}}_- & = & -2 \, d {\bar \gamma_I} (z,{\bar z} )\, \wedge \,
  \omega^I\
  \end{array}
\label{hpmposiz}
\end{equation}
where $\gamma_I (z,{\bar z} )$ is a complex field depending only
on the  coordinates $z,\bar z$ of $\mathbb{R}^2$, while $\omega^I$
($I=1,\dots,k$) constitute a basis for the space of square
integrable, anti-self-dual, harmonic forms on the ALE manifold.
\par
As it is well known \footnote{See for instance \cite{Anselmi:1994sm}
for an early summary of ALE geometry in relation with superstrings
and conformal field theories. This relation was developed in
\cite{Douglas:1996sw,Johnson:1997py} and is of primary
relevance in connection with D-branes.} a smooth ALE manifold,
arising from the resolution of a $\mathbb{C}^2/\Gamma$ singularity,
where $\Gamma \subset \mathrm{SU(2)}$ is a discrete Kleinian
group, has Hirzebruch signature:
\begin{equation}
  \tau =\left( \# \mbox{of conjugacy classes of $\Gamma$}\right) - 1
       = \mbox{rank of } \,\mathbb{G}(\Gamma)
\label{Hirze}
\end{equation}
In the above formula $\mathbb{G}(\Gamma)$ is the simply laced Lie
algebra corresponding to $\Gamma$ in the ADE classification of
Kleinian groups, based on the Mac Kay correspondence
\cite{mackay}. As a result of eq.(\ref{Hirze}) the ALE manifold
that is HyperK\"ahler admits a triplet of self-dual $2$-forms that
are non-integrable and exactly $\tau$ integrable anti-self-dual
harmonic $2$-forms. For these latter one can choose a basis
$\omega^I$ that is dual to the integral homology basis of
$2$-cycles $\Sigma_I$ whose intersection matrix is the Cartan
matrix $\mathcal{C}$ of $\mathbb{G}(\Gamma)$.  Explicitly we can
write:
\begin{equation}
\begin{array}{ccccccccccc}
\int_{\Sigma_K} \,\omega^I&=&\delta^I_K &;&
\int_{ALE} \omega^I\wedge \omega^J &=&-\,{\cal C}^{-1\, IJ} &;&
\omega^I\wedge \omega^J &=& -\,\Delta^{IJ}(y)\,\Omega_{ALE}\\
\end{array}
\label{omega}
\end{equation}
where ${\cal C}^{IJ}$ is the Cartan matrix of the corresponding
(non-extended) ADE Dynkin diagram and $\Delta^{IJ}(y)$ is a
positive definite matrix whose entries are functions of the ALE
space coordinates $y$'s. The anti-self-duality of the $\omega^I$
guarantees that $\int \omega^I\wedge\star \omega^J$ is positive
definite.
If we insert our  ans\"atze  into
the scalar field equations (\ref{NSscalapr}, \ref{RRscalapr}) we
obtain ${\mathcal{H}}_+ \, \wedge \, \star{\mathcal{H}}_+=0 $
which  in turn implies that
$\partial_z\gamma_I\,\bar{\partial}_{\bar{z}}\gamma_J=0$. This
equation is solved by choosing $\gamma_I$ to be holomorphic:
$\bar{\partial} \gamma_I=0 $ where $\partial = dz \, \frac{\partial}{\partial
  z} $, $ {\bar \partial} = d{\bar z} \, \frac{\partial}{\partial
  {\bar z}}$.
Next we consider the self-dual $5$-form $F_{[5]}^{RR}$ which,
because of its definition, must satisfy the following Bianchi
identity:
$d \, F_{[5]}^{RR} = {\rm i} \, \ft 1 8 \,{ \mathcal{H}}_+
\, \wedge \, { \mathcal{H}}_- $.
Our ansatz for $F_{[5]}^{RR}$ is the following: (
$\Omega$ are the volume forms)
\begin{equation}
F_{[5]}^{RR}  =  \alpha \left( U + \star U \right)  \quad ; \quad
U  =   d \left( H^{-1} \, \Omega_{\mathbb{R}^{1,3}} \right)
\label{f5ansaz}
\end{equation}
where $\alpha$ is a constant to be determined later. By
construction $F_{[5]}^{RR}$ is self-dual and its equation of
motion is trivially satisfied. What is not guaranteed is that also
the $5$--form Bianchi identity is fulfilled. Imposing it,
results into a differential equation for the function $H(z,\bar z,
y)$:
\begin{equation}
  \left( \square_{\mathbb{R}^2} + \square_{ALE} \right)  H = - \frac
  {1}{\alpha} \partial_z \gamma_I \,
  \partial_{\bar z}
  {\bar \gamma}_J \, \Delta^{IJ}(y)
\label{maindiffe}
\end{equation}
This is the main differential equation the entire construction of
our sought for 3-brane solution can be reduced to. The parameter $\alpha$ is determined by Einstein's
equation and fixed to $\alpha=1$.
\par
The field equation for the complex
three-form, namely eq.s (\ref{3formNS}) and  (\ref{3formRR})  reduces to:
$
   {\bar \partial} \wedge \partial \gamma_I = 0 \, \Rightarrow \,
   \square_{\mathbb{R}^2} \gamma_I = 0
$.
This equation has  to be appropriately interpreted. It says that
$\gamma_I$ are harmonic functions in two-dimensions as the real
and imaginary parts of any holomorphic function $\gamma_I (z)$
certainly are. The bulk equations do not impose any additional
constraint besides this condition of holomorphicity. However, in
presence of a boundary action for the $3$ brane, the equation
will be modified into:
\begin{equation}
  \square_{\mathbb{R}^2} \gamma_I = j_I(z)
\label{consorgo}
\end{equation}
$j_I(z)$ being a source term, typically a delta function. In this case $\gamma_I$
is fixed as:
$
  \gamma_I (z) = \int \, G_{\mathbb{R}^2} (z,z^\prime)\,  j_I(z^\prime) \, dz^\prime
$
where $G_{\mathbb{R}^2} (z,z^\prime)$ is the Green function of the
$\mathbb{R}^2$ Laplacian in complex coordinates and turns out to
be proportional to $\log z$.
\subsection{Proof of bulk supersymmetry}
\label{susyprova}
As usual, in order to investigate the
supersymmetry properties of the bosonic solution we have found it
suffices to consider the supersymmetry transformation of the
fermionic fields (the gravitino and the dilatino) and impose that, for a
Killing spinor, they vanish identically on the chosen background.
By using the formulation of \cite{igorleo}, one easily gets:
\begin{eqnarray}
\delta\psi_M & = & \mathcal{D}_M\chi + \ft 1 {16} \, {\rm i}
\Gamma^{A_1 \dots A_5}  \,
F_{A_1 \dots A_5}\, V^{B}_M \, \Gamma_{B} \, \chi \,\nonumber\\
&& +\ft 1 {32} \left( - \Gamma_{BA_1\dots A_3} \, V^{B}_M+ 9
\Gamma^{A_1 A_2} V^{A_3}_M \right) \,
{\mathcal{H}}_{+\vert A_1 A_2 A_3} \, \chi^\star
\nonumber\\
\delta \, \lambda & = & -{\rm i} \, \ft 1 8 \, \Gamma^{A_1 A_2
A_3} \,{\mathcal{H}}_{+\vert A_1 A_2 A_3}\, \chi
\label{susyvaria}
\end{eqnarray}
where the supersymmetry parameter $\chi$ is a complex
ten-dimensional Weyl spinor,
$
  \Gamma_{11} \, \chi = \chi
$
and where we have already used the information that on our background
the dilaton and the Ramond scalar vanish.
To analyze supersymmetry on such a background the appropriate gamma
matrix basis $\left(  \left\{ \Gamma^{A} \, , \, \Gamma^{B} \right\}  =  \eta^{AB}\right) $ is the following:
\begin{eqnarray}
\Gamma^{A}& =& \left\{ \begin{array}{rcl}
  \Gamma^a & = & \gamma^a \,\otimes {\bf 1} \\
  \Gamma^i & = & \gamma_5 \, \otimes \, \tau^i \,
\end{array} \right. \label{grossgamma}
\end{eqnarray}
where
$
\left\{ \gamma^a \, , \, \gamma^b\right\}   =  2\, \eta^{ab}  $
, $ \gamma_5  =  - {\rm i} \gamma^0 \gamma^1 \gamma^2  \gamma^3
$, $
\left\{ \tau^i \, , \, \tau^j\right\}   =  2\, \eta^{ij} = - 2
\, \delta^{ij} $ , $ \tau_\bullet = - \tau^4 \tau^5  \tau^6
\tau^7  \tau^8 \tau^9$
are the gamma matrices in Lorentzian four space and on the six
dimensional manifold $M_6$ respectively.  Then the
$\tau^i$ matrices are further decomposed with respect to the submanifolds $ \mathbb{R}^2$ and
$\mathrm{ALE}$ as it follows:
\begin{equation}
  \tau^i =\left\{ \begin{array}{rcl}
    \tau^\alpha & = & {\rm i} \sigma^{\alpha} \, \otimes \, {\bf 1} \\
    \tau_u & = & {\rm i} \sigma^3 \, \otimes \, \overline{\gamma} _u\
  \end{array} \right.
\label{tausplitte}
\end{equation}
where $\sigma^\alpha = \sigma^{1,2}$ and $\sigma^3$ are the
standard Pauli matrices while $
 \left\{  \overline{\gamma} _u \, , \,  \overline{\gamma} _v \right\}
 = \delta_{uv}$
are $4 \times 4 $ hermitian matrices forming an Euclidean realization of the four-dimensional Clifford algebra.
Writing the $32$-component spinor $\chi$ as a tensor product:
\begin{equation}
  \chi = \varepsilon \,  \otimes \, \eta
\label{tensprodspin}
\end{equation}
of a $4$-component spinor $\varepsilon$,
related to the $3$-brane world volume
with an $8$-component spinor $\eta$ related to the transverse
manifold $M_6$, the transformations of the gravitino and dilatino
(\ref{susyvaria}) vanish if:
\begin{equation}
\begin{array}{lclclcl}
  \partial_a \, \varepsilon =0  &;& \gamma_5 \, \varepsilon =
   \varepsilon &; & \eta = H^{-1/8}(z, \bar z, y) \, \theta \, \otimes \,
   \xi & ; &
  0 = \widehat{\mathcal{D}} \xi \\
   \overline{\gamma}_5 \xi_- = - \xi_- & ; & \sigma_3 \, \theta \, = -
  \theta & ; &
  0=\partial _z \theta &; & 0=\partial_{\bar z} \theta
\end{array}
\label{condiziones}
\end{equation}
The specific geometric properties of the
ALE manifold play at this point an essential role. The integrability
condition for covariantly constant spinors $\xi$ is, as usual given
by
$
  - \ft 1 4 \, R^{uv} \, \overline{\gamma}_{uv} \, \xi =0
$
where $R^{uv} \, \overline{\gamma}_{uv}$ is the curvature $2$-form
of the ALE manifold. This latter is HyperK\"ahler and as such it
has a triplet of {\sl covariantly constant self-dual} $2$-forms
$\Omega^x$, $(x=1,2,3)$ whose intrinsic components satisfy the
quaternionic algebra. This implies that the holonomy of the
manifold is $\mathrm{SU(2)}_L$ rather than $\mathrm{SO(4)} =
\mathrm{SU(2)_L }\times \mathrm{SU(2)_R}$ and that the curvature
$2$-form is {\sl anti-self-dual}. This follows from the
integrability condition for the covariant constancy of the
self-dual HyperK\"ahler $2$-forms. On the other hand, from the
Hirzebruch signature $\tau$ of the ALE manifold it follows that
there are exactly $\tau$ normalizable anti-self-dual forms
$\omega_I$. From the trivial gamma matrix identity
$
  \overline{\gamma}_5 \, \overline{\gamma}_{uv} = - \ft 1 2 \,
  \epsilon_{uv st} \, \overline{\gamma}^{st}
$
follows that the two chirality eigenspaces:
$
  \overline{\gamma}_5 \xi_\pm = \pm \xi_\pm
$
are respectively annihilated by the contraction of
$\overline{\gamma}_{uv}$ with any self-dual or anti self-dual
$2$-form. Therefore antichiral spinors
$
  \xi= \xi_-
$
satisfy the integrability condition  automatically .
Once this condition is
fulfilled, the equation $\widehat{D}_a\xi=0 $ can be integrated yielding
two linear independent solutions $\xi_-^{1,2}$ that span the
irreducible representation $(0,1/2)$ of $\mathrm{SO(4)}$. The
other irreducible representation $(1/2,0)$ corresponds to spinors
that are not Killing and do not generate supersymmetries preserved
by the background. In conclusion we have $2$ Killing spinors
generating an $\mathcal{N}=2$ supersymmetry on the world volume.
In other words the bosonic background we have constructed
corresponds to a BPS state preserving a total of $2 \times 4 = 8$
supercharges.
\section{The Eguchi-Hanson case}
\label{Ehcase}
As showed above, the complete
integration of the supergravity field equations is reduced
to the solution of a single differential equation,
namely eq.
(\ref{maindiffe}). It is worth to investigate
the properties of such an equation, choosing the simplest instance
of an  ALE manifold, namely the Eguchi-Hanson space
\cite{Eguchi:1978xp}, which in the ADE classification corresponds to
$A_1 \sim \mathbb{Z}_2$. In paper \cite{paperus}
a complete and detailed mathematical analysis of such
a case was given. In this talk I summarize
the main results of such an analysis.
It should be stressed that, whereas the results
of the previous
sections were entirely based
on type IIB supergravity, in what follows
 we have to make some  reasonable
 assumptions on the nature of the microscopic theory,
namely the structure of the source terms needed
to fix the boundary
conditions.
\par
The Eguchi-Hanson metric has the form:
\begin{equation}
ds^2_{EH}= g(r)^{-2}\,dr^2+ r^2\,\left(\sigma_x^2+
\sigma_y^2\right)+r^2\,g(r)^2\,
\sigma_z^2 \quad ; \quad
g(r)^2=1-\left(\frac{a}{r}\right)^4 \label{eguchi}
\end{equation}
where $r\ge a$, $\theta\in \left[0,\,\pi\right[$ and $\psi$, $\phi
\in \left[0,\,2\pi\right[$. The $1$-forms $\sigma^i=\left\{
\sigma^x, \, \sigma^y ,\, \sigma^z \right \}$  satisfy, by definition the Maurer Cartan equations
of $SO(3)$ and are defined on the three sphere.\par This space has a unique
homology 2-cycle $\Sigma$ located at $r=a$ and spanned by the
coordinates $\theta$, $\phi$. The anti-self dual form $\omega$
fulfilling eqs. (\ref{omega}) is:
\begin{eqnarray}
\omega &=&\frac{a^2}{2\,\pi}\,d\left(\frac{\sigma_z}{r^2}\right)
\end{eqnarray}
The function $\Delta(y)$ defined in (\ref{omega}) is explicitly
evaluated to be:
\begin{eqnarray}
\Delta(y)&=&\frac{2\, a^4}{\pi^2\,r^8}\label{Delta}
\end{eqnarray}

The equation for $H$ in the Eguchi-Hanson case can then
be easily obtained from
the general expression in eq.(\ref{maindiffe}). As it is usually the case,
we make a spherically symmetric ansatz, compatible with the background at hand:
for the case we shall be interested in, the
coefficients of the equation for $H$ depend only on the
radial coordinates $\rho=\sqrt{z\bar{z}}$ and $r$ on
$\mathbb{R}^2$ and on the Eguchi-Hanson space respectively,
and we can then assume the same property for $H$ to hold and write it as
$H(\rho,\,r)$\footnote{We shall consider the case in which
$\gamma(z)=\log{(z)}$. In a more general situation
$\vert\partial\gamma(z)\vert^2$ could also depend on the angular
coordinate on $\mathbb{R}^2$. In this case the function $H$ would
have an angular dependence in $\mathbb{R}^2$ as well.}.
The equation for $H$ can then be written as:
\begin{eqnarray}
\left(\partial_4^2+\partial_5^2\right)\,H+\frac{1}{r^3}\partial_r\left(r^3\,g(r)^2\,\partial_r
H\right)&=& -\frac{2\,
a^4}{\pi^2\,r^8}\,|\partial_z\,\gamma|^2+{\cal S}(\rho,\,r)
\label{maindifEH}
\end{eqnarray}
where ${\cal S}(\rho,\,r)$ is a source term for the $3$-brane
charge for which we also make a spherical ansatz. Differently from the first
term on the right hand side, ${\cal S}(\rho,\,r)$ is not deduced from the
dynamics in the bulk. In the
present analysis its presence should be intended
only for the sake of fixing boundary conditions
near the cycle.
\par
The best technique to study eq.(\ref{maindifEH})
is that
of performing a Bessel--Fourier transform.
For convenience we choose to study the
equation in the Fourier transform
$\widetilde{H}(\mu,r)$ of $H(\rho,r) - 1$ instead of
$H(\rho,r)$.  In this way the boundary
conditions at infinity: $ H(\rho,r) \, \rightarrow\,1$
is automatically implemented if $\widetilde{H}(\mu,r)\, \rightarrow\,0$.
The relation between $\widetilde{H}(\mu,r)$
and $H(\rho,r)$ is given by:
$
  \widetilde{H}(\mu,r) =\frac{1}{2\pi}\int_0^{\infty}
 d\rho \,\rho\, J_0(\mu\rho) \, \left(H(\rho,r)-1\right)
$.
The equation for $\widetilde{H}(\mu,\,r)$ has the following form:
\begin{equation}
 \left[  \frac{1}{r^3} \frac{d}{dr} \, \left( r^3 \, g(r)^2 \, \frac{d}{dr}
  \right)  - \mu^2 \,\right]  \, \widetilde{H}(\mu,\,r) = J(\mu,\,r)
\label{formalform}
\end{equation}
where we have defined the source function:
$
  J(\mu,\,r)  \, \equiv \, -\frac{2 a^4}{\pi^2 \,
  r^8}\, j(\mu)+ S(\mu,\,r)
$. The symbol
$j(\mu)$ denotes the Fourier transform of
$\vert\partial\gamma(z)\vert^2$ while  $S(\mu,\,r)$   is the
transform of the source term ${\cal S}(\rho,\,r)$   for the
$3$-brane charge.
\par
It is known from the standard theory of differential
equations that the general integral $\widetilde{H}(\mu,\,r)$ of
eq. (\ref{formalform}) has the form:
\begin{equation}
\widetilde{H}(\mu,\,r) = \beta_1\, u_1(\mu,\,r) + \beta_2\, u_2(\mu,\,r) + u_{n-h}(\mu,\,r)
\label{genintegra}
\end{equation}
where $u_{1,2}(\mu,\,r)$ are two independent solutions of the
homogeneous equation associated with (\ref{formalform}).
\subsection{The $3$-brane charge and the
physical boundary
conditions}
\label{d3branacargo}
The physical boundary conditions for the $D3$--brane
solution are imposed by selecting the asymptotic
behaviour of the warp factor $H$ near infinity $r \to
\infty$ and near the cycle at $r \to a$.
In order to perform such an analysis we just need
to consider the two asymptotic limits $r \to
\infty$ and $r \to a$ of the Eguchi Hanson metric.
As for the limit $r \to \infty$ the limit is clear, the Eguchi Hanson
metric approaches the flat metric and this is just
what ALE means,
namely asymptotically locally Euclidean.
The near cycle limit
of the same metric metric (\ref{eguchi}) is
exposed by performing the following change of variable:
$
 r\rightarrow v  =
    \frac{{\sqrt{ r^4-a^4 }}}{2\,r}
$.
By expanding the
metric (\ref{eguchi}) in power series of
$v$ at $v=0$ which corresponds to $(r \to a)$,
we obtain:
\begin{eqnarray}
ds^2_{EH} & \simeq_{v \to 0} & dv^2 +v^2\,d\psi^2
 + \frac{a^2}{4} \, \left ( d\theta^2
   \, +  \,
     d\phi^2\,\sin^2 \theta  \right ) +\mathcal{O}\left(v^2\right)
\label{ubipicchio}
\end{eqnarray}
showing that near the homology cycle the Eguchi-Hanson metric
approximates that of a manifold $ \mathbb{R}^2 \, \times \,
\mathrm{S}^2$.
\par
 Equipped these results, let us now consider the case of a $3$-brane with vanishing 3-form
placed either at the origin of $ \mathbb{R}^6$ in the orbifold
case where the transverse space is $M_6 =\mathbb{R}^2 \times
\mathbb{R}^4/\Gamma$  or at the homology cycle $r=a$ in the case
where the transverse space is $M_6 =\mathbb{R}^2 \times
\mathrm{ALE_{EH}}$. Naming $Q$ the charge of such a brane, the
expected behavior of the $H$ function near the brane is, in the
two cases, the following one:
\begin{equation}
H \simeq \frac{Q}{(x_4^2 + x_5^2 + \dots+x_9^2)^2}
\quad\mbox{for $ \mathbb{R}^6$; }\quad
H \simeq  \frac{Q/{2 a^2 }}{\, (\rho^2 + v^2)}\quad
\mbox{for $ \mathbb{R}^2 \times$ ALE }
\label{rpowers}
\end{equation}
The first of eq.s (\ref{rpowers}) is obvious. The second is due to
the discussion of the previous section. Near the cycle
the Laplacian on $M_6$ becomes
$
  \square_{M_6} \sim \square_{\mathbb{R}^4} +\square_{S^2}
$
and since $S^2$ is a compact, positive curvature manifold there
are no zero modes of $\square_{S^2}$ except the constant.
Therefore the non trivial part of $H$ behaves as a harmonic
function on $\mathbb{R}^4$, namely the second of eq.s
(\ref{rpowers}). The  scale factor $1/(a^2)$ appearing there
is understood as follows. If $Q$ is
the total charge perceived at infinity, the density of charge on
the homology $2$-sphere of radius $\frac a 2$ is:
$
  q=\frac{Q}{a^2 \pi}
$.
Hence what appears as charge in the near cycle $\mathbb{R}^4$
plane is $q$ rather than $Q$. Finally the $\pi/2$ factor that  is
needed to match the second with the first of equations
(\ref{rpowers}) is just a matter of convenient normalization. Let
us now perform the Fourier-Bessel transform of eq.s
(\ref{rpowers}). We obtain
\begin{eqnarray}
\frac{1}{2\pi} \, \int_0^\infty  \, J_0(\mu\,\rho) \, \rho \,
\frac{Q}{(r^2+\rho^2)^2} \, d\rho & = & \frac{Q}{4 \pi} \,
\frac{\mu}{r} \, K_1\left (\mu\,  r\right) \nonumber\\
\frac{1}{2\pi} \, \int_0^\infty  \, J_0(\mu\,\rho) \, \rho \,
\frac{Q}{2 a^2 \, \left (v^2+\rho^2 \right )} d\rho & = &
  \frac{Q}{4 \pi} \,
\frac{1}{a^2 }  \, K_0\left (\mu \, v\right) \label{Qbrancargo}
\end{eqnarray}
The important conclusion implied by the above analysis is that we have
obtained the physically appropriate boundary conditions for the
function $\widetilde{H}(\mu,r)$. In both the orbifold or smooth
ALE case, in the limits $r\rightarrow 0$ and $r\rightarrow a$
respectively, we have:
$
  \widetilde{H}(\mu,r)  \sim
  \frac{Q}{4 \pi}  {u}^{div}_1(\mu,\,r)
  + \mbox{reg}
$
where $Q$ is the $3$-brane charge and ${u}^{div}_1(\mu,\,r)$
denotes the divergent part of that of the two solutions  $u_{1,2}(a,\,\mu,\,r)$
of the homogeneous Laplacian equation that is divergent in the limit.
This condition, together with the
boundary condition at infinity,  fixes the coefficients
$\beta_1,\,\beta_2$ in eq. (\ref{genintegra}) to be:
$
  \beta_1 = \frac{Q}{4 \pi} \, ; \, \beta_2 = 0
$.
The appropriate  source for such boundary conditions is:
$
S(\mu,\,r) = - \frac{Q}{2 \pi} \, \frac{\delta(r-a)}{r^3}
$.
 From the physical viewpoint this source term comes from the world-volume action of the
$D3$-brane.

\subsection{Reduction to the confluent Heun equation}
\label{heunsoft}
As we have seen in previous sections a convenient
approach to the solution of  equation (\ref{maindifEH}) relies on
the partial Fourier transform leading to
the new eq. (\ref{formalform}),  for the function
$\widetilde{H}(\mu,r)$. In \cite{paperus} it was shown that complete
equation reduces to a confluent form of the Heun equation. This is
obtained by parameterizing the radial
direction in the Eguchi-Hanson space through the new variable
$w=(a^2-r^2)/(2 a^2)$ so that the main differential equation
(\ref{formalform}) can be rewritten in the
following form:
\begin{eqnarray}
w^2\,\frac{d^2 \widetilde{H}}{dw^2}+w\,P_1(w)\,\frac{d
\widetilde{H}}{dw}+ P_2(\mu,\,w)\, \widetilde{H}&=&
\frac{h(\mu)\, w}{(1-w)(1-2w)^3}\nonumber\\
\widetilde{H}(\mu,w)&\stackrel{ w\rightarrow
-\infty}{\longrightarrow}& 0
 \label{heun}
\end{eqnarray}
where
$
P_1(w) = \frac{2w-1}{w-1}$, $
P_2(\mu,\,w) = k\,w\,\left(\frac{2w-1}{w-1}\right) $, $
h(\mu) = \frac{1}{2\pi^2\, a^{2}}\,j(\mu)$ =$ \frac{1}{2 \pi^2
a^2}\int_0^\infty\, \frac{d^2 x}{(2\pi)^2}\,\vert\partial_z
\gamma(z)\vert^2\,e^{{\rm i}\vec{p}\cdot \vec{x}} $,  $ k =
\frac{\mu^2 a^2}{4}$.
In the source term we omit the part proportional to
$S(\mu,\,r)$ since, as just explained  its
effect amounts to a determination of the relative coefficients of
the two independent solutions $u_{1,2}$ of the homogeneous
equation.
\par
The power series solution of eq.(\ref{heun}) explicitly discussed in
\cite{paperus} is a smooth interpolation between the two asymptotic
behaviours of the warp factor that we recall here in complete form.
Far away from the cycle we have:
\begin{equation} \left(\rho, r\gg a\right):\quad H(\rho,\,r)\sim
\frac{Q}{(\rho^2+r^2)^2}\left(1 + \mathrm{log~terms} \right)
\end{equation}
while in the region near the cycle, combining the asymptotic
behaviors of the homogeneous and inhomogeneous solutions, we find
\begin{equation}
\label{Hregcy}
\left(\matrix{ \rho< a\cr r\sim a }\right):\quad
H(\rho,\,r)\sim \frac{Q}{2a^2}\frac{1}{\rho^2+v^2} -
\frac{1}{2\pi^2\,a^4}\,K\,\log{\left(\frac{\rho}{\rho_0}\right)}^2
\end{equation}
where $K$ is a constant, completely free at the level of the bulk supergravity analysis. The
corrections to the above behavior can be systematically deduced from the power
series expansion of the solution described in \cite{paperus}.
Looking at eq.(\ref{Hregcy}), it is clear that there could be a
value of $r,\rho$ for which $H=0$, this being an indication for the
presence of a naked singularity of the repulson type \cite{kal}. This
singularity should be removed, somehow. In most non-conformal
$\mathcal{N}=2$ versions of the gauge/gravity correspondence this
singularity has been shown to be excised by the so-called
enhan\c{c}on mechanism \cite{enhanc}. This is the case, for instance, of
fractional branes on orbifolds, \cite{Bertolini:2000dk},
\cite{Anselmi:1994sm}. The value of the
enhan\c{c}on corresponds to $\rho=\rho_e$, the scale where the scalar
field $\gamma$ vanishes. This in general turns out to be the scale where the
dual gauge theory becomes strongly coupled and new light degrees of freedom
are expected to become relevant, both at the gauge theory (where instanton
effects become important) and at the supergravity level (where
tensionless strings occur). For all this analysis to work, it is
important that $\rho_e$ is bigger than the scale at which the repulson
occurs. In fact, when this is the case, the region where supergravity
is reliable, namely $\rho>\rho_e$, is free of any singularity. In
order to see if this happens also in our case and if the cut-off
$\rho_e$ has in fact the expected meaning, one should have a full
control on the world-volume action of the source. However
the  solution I have presented   differs  from that of fractional branes on
singular space because of the improved behaviour of the warp factor $H$ on the
plane $\mathbb{R}^2$ where it is non-singular, while the solution of the field
$\gamma$, which is responsible for the enhan\c{c}on mechanism,
has essentially the same structure.  Hence this bulk solution is reliable, well-defined
and singularity free for $\rho>\rho_e$ where the leading order
near cycle behaviour is well described by the first term in
eq. (\ref{Hregcy}).
\iffigs
\begin{figure}
\begin{center}
\epsfxsize =12cm
{\epsffile{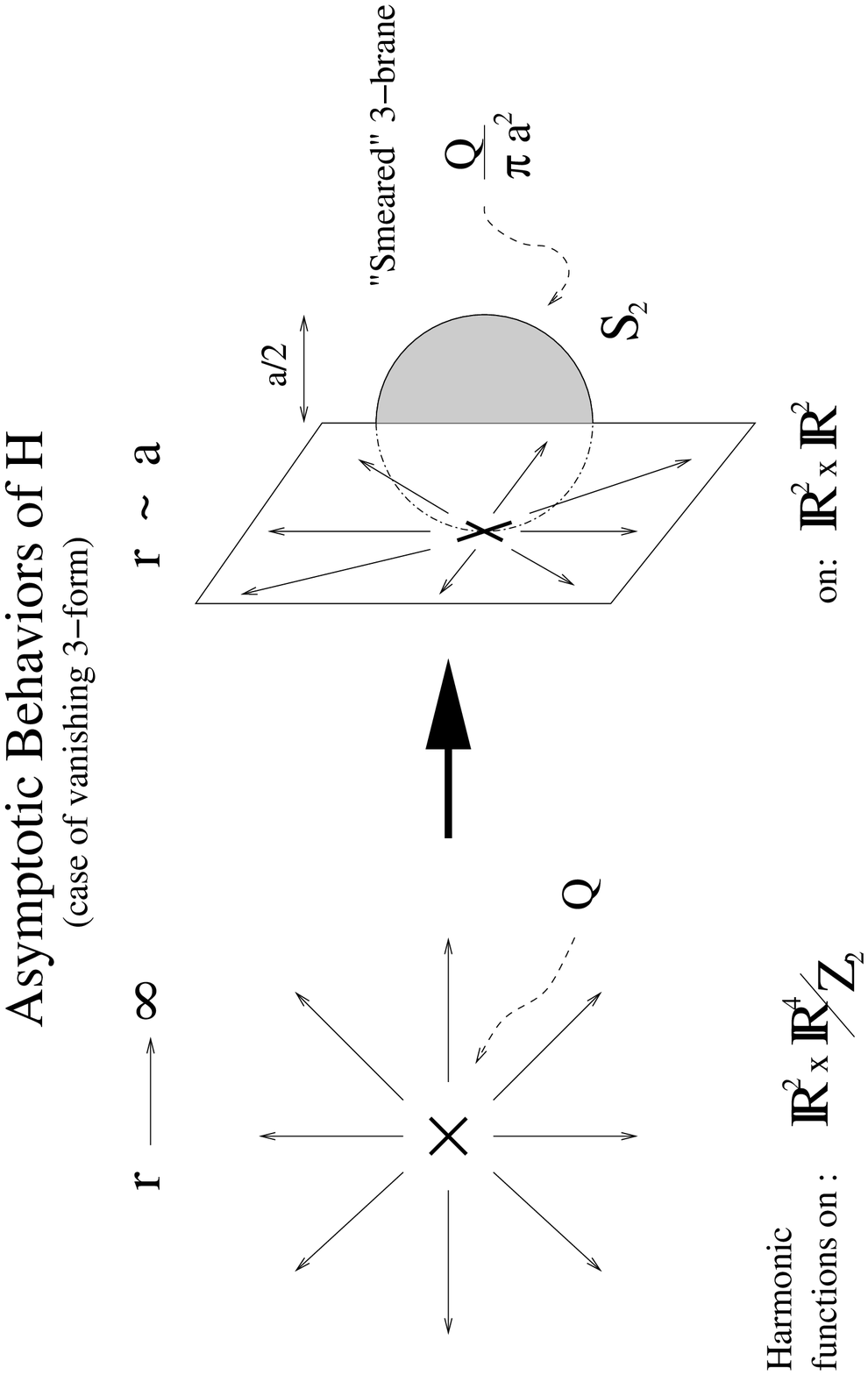}
\caption{\label{figu1} }
}
\hskip 2cm
\unitlength=1.1mm
\end{center}
\end{figure}
\fi
\par
\section{Final considerations on the metric and open problems}
\label{conc}
Introducing the variable $R = \sqrt{(\rho^2 + v^2)}$ that measures
the radial coordinate of $\mathbb{R}^4$ in units of $q$ we find
that near the cycle we have $H \sim q/R^2$, yielding
\begin{equation}
ds^2 \sim \frac{R}{\sqrt{q}}(-\eta_{\mu\nu}dx^\mu dx^\nu) +
 \frac{\sqrt{q}}{R} dR^2+
\sqrt{q} R ds^2_{\mathbf{S}^3} + \frac{\sqrt{q}}{R}
\frac{a^2}{4} ds^2_{\mathbf{S}^2}
\end{equation}
where $ds^2_{\mathbf{S}^2}$ is the metric of the two cycle of
Eguchi-Hanson and $ds^2_{\mathbf{S}^3}$ is the metric of the
three-sphere at fixed $R$ in $\mathbb{R}^4$. An interesting point
is that the above result holds even in the absence of the
three-form flux and shows that conformal invariance of the dual
theory is always broken since the metric in $R$ and $x^\mu$ is
no longer of anti de Sitter type. Obviously we cannot take the
limit $R \to 0$ since this would imply going to large
curvatures, and in particular crossing the enhan\c{c}on radius.
However, the result suggests that a possible interpretation for
the parameter $a$ in the Eguchi-Hanson metric is that of a Fayet
Iliopoulos term, breaking conformal invariance in the infrared, in
accord with previous work on the subject.
\par
The situation can be summarized as follows. In the absence of flux the
exact supersymmetric $D3$-brane solution that we have found interpolates
between a standard ten dimensional $D3$-brane solution at the singularity of
the metric cone on $S^5$, i.e. the standard $ \mathbb{R}^6$ manifold, and
a $D3$-brane solution of an effective $8$-dimensional supergravity. The interpolation
mechanism is described in fig. \ref{figu1}. As explained
in the literature \cite{popino1} (for a review see \cite{parilez}) we
can always consider sphere reductions of all supergravity theories and in
particular an $S^2$-reduction of type IIB supergravity.
This yields an effective
$8$-dimensional supergravity that has $3$-brane solutions.
In this case however,
there is a coupling to an effective dilaton  that emerges as the conformal
factor of the metric in the dimensional reduction. Hence the $D3$-brane solution
of the $8$-dimensional supergravity is no longer conformal and we can follow
the prescription of Townsend et al \cite{Boonstra:1999mp} by making the transition
to a {\it dual frame} where the metric factorizes into the product of a
$3$-sphere
metric times a domain wall solution of an effective five-dimensional
supergravity
theory.
\par
Investigating the relation with the effective $8$--dimensional
supergravity and the properties of the dual gauge theory on the world
volume are the challenging open problems posed by the type IIB
solution I have described in this talk.
\vskip 1.cm
\noindent
{\large {\bf Acknowledgments}} It is my pleasure to thank my
collaborator and friend Mario Trigiante for help in preparing this
talk and all the coauthors of paper \cite{paperus} on which
what I told is entirely based.
\vskip 0.5cm
\noindent


\begin{thebibliography}{99}
\bibitem{paperus} M. Bertolini, V.L. Campos, G. Ferretti, P. Fr\'e,
P. Salomonson, M. Trigiante. {\it Supersymmetric 3--branes on smooth
ALE manifolds with flux} hep-th/0106186, to appear on Nucl. Phys.
%
\bibitem{mal}
J. Maldacena, \emph{The large N limit of superconformal field
theories and supergravity}, Adv. Theor. Math. Phys. {\bf 2} (1998)
231, {\tt hep-th/9711200}.
%
\bibitem{Klebanov:2000rd}
I.~R.~Klebanov and N.~A.~Nekrasov, \emph{Gravity duals of
fractional branes and logarithmic RG flow}, Nucl.\ Phys. {\bf
B574} (2000) 263, {\tt hep-th/9911096}.
%
\bibitem{KLEBA2} I.R. Klebanov and A.A. Tseytlin,
\emph{Gravity duals of supersymmetric SU(N)*SU(N+M) gauge
theories}, Nucl. Phys. {\bf B578} (2000) 123, {\tt
hep-th/0002159}.
%
\bibitem{TATAR}K. Oh and R. Tatar,
\emph{Renormalization group flows on D3 branes at an orbifolded
conifold}, JHEP {\bf 05} (2000) 030, {\tt hep-th/0003183}.
%
\bibitem{KLEBA3} I.R. Klebanov and M.J. Strassler,
\emph{Supergravity and a confining gauge theory: duality cascades
and $\chi$B-resolution of naked singularities}, JHEP {\bf 08}
(2000) 052, {\tt hep-th/0007191}.
%
\bibitem{TSEYTLIN1} L.A. Pando Zayas and A.A. Tseytlin,
{\em 3-branes on resolved conifold}, {\tt hep-th/0010088}.
%
\bibitem{Bertolini:2000dk}
M.~Bertolini, P.~Di~Vecchia, M.~Frau, A.~Lerda, R.~Marotta and
I.~Pesando, \emph{Fractional D-branes and their gauge duals},
JHEP {\bf 02} (2001) 014, {\tt hep-th/0011077}.
%
\bibitem{Polchinski:2000mx}
J.~Polchinski, \emph{N = 2 gauge-gravity duals}, {\tt
hep-th/0011193}.
%
\bibitem{anto}
M. Frau, A. Liccardo and R. Musto, \emph{The geometry of
fractional branes}, Nucl. Phys. {\bf B602} (2001) 39, {\tt
hep-th/0012035}.
%
\bibitem{Aharony}
O.~Aharony, \emph{A note on the holographic interpretation of
string theory backgrounds with varying flux}, JHEP {\bf 03}
(2001) 012, {\tt hep-th/0101013}.
%
\bibitem{Petrini:2001fk}
M.~Petrini, R.~Russo and A.~Zaffaroni, \emph{N = 2 gauge theories
and systems with fractional branes}, {\tt hep-th/0104026}.
%
\bibitem{Aharony:1998xz}
O.~Aharony, A.~Fayyazuddin and J.~Maldacena, \emph{The large N
limit of N = 2,1 field theories from three-branes in  F-theory},
JHEP {\bf 07} (1998) 013, {\tt hep-th/9806159}.
%
\bibitem{Brinne:2000fh}
B.~Brinne, A.~Fayyazuddin, S.~Mukhopadhyay and D.~J.~Smith,
\emph{Supergravity M5-branes wrapped on Riemann surfaces and their
QFT duals}, JHEP {\bf 12} (2000) 013, {\tt hep-th/0009047}.
%
\bibitem{Herzog:2001rz}
C.~P.~Herzog and I.~R.~Klebanov, \emph{Gravity duals of fractional
branes in various dimensions}, Phys.\ Rev. {\bf D63} (2001)
126005, {\tt hep-th/0101020}.
%
\bibitem{marco} M. Bill\`o, L. Gallot and A. Liccardo, {\em Classical geometry and gauge duals for fractional
branes on ALE spaces}, {\tt hep-th/0105258}.
%
\bibitem{Anselmi:1994sm}
D.~Anselmi, M.~Billo, P.~Fre, L.~Girardello and A.~Zaffaroni,
\emph{ALE manifolds and conformal field theories}, Int.\ J.\ Mod.\
Phys. {\bf A9} (1994) 3007, {\tt hep-th/9304135}.
%
\bibitem{Douglas:1996sw}
M.~R.~Douglas and G.~Moore, \emph{D-branes, Quivers, and ALE
Instantons}, {\tt hep-th/9603167}.
%
\bibitem{Johnson:1997py}
C.~V.~Johnson and R.~C.~Myers, \emph{Aspects of type IIB theory on
ALE spaces}, Phys.\ Rev.\  {\bf D55} (1997) 6382, {\tt
hep-th/9610140}.
%
\bibitem{Douglas:1997xg}
M.~R.~Douglas, \emph{Enhanced gauge symmetry in M(atrix) theory},
JHEP {\bf 07} (1997) 004, {\tt hep-th/9612126}.
%
\bibitem{Diaconescu:1998br}
D.~Diaconescu, M.~R.~Douglas and J.~Gomis, \emph{Fractional branes
and wrapped branes}, JHEP {\bf 02} (1998) 013, {\tt
hep-th/9712230}.
%
\bibitem{kal} R. Kallosh and A. Linde, Phys. Rev. {\bf D52} (1995) 7137,
{\tt hep-th/9507022}.
%
\bibitem{bflux}
P. Aspinwall, \emph{Enhanced gauge symmetries and K3 surfaces},
Phys. Lett. \textbf{B357} (1995) 329, {\tt hep-th/9507012}; W.
Nahm and K, Wendland, \emph{A hiker's guide to K3: Aspects of N =
(4,4) superconformal field theory  with central charge c = 6},
Commun. Math. Phys. \textbf{216} (2001) 85, {\tt hep-th/9912067}.
%
\bibitem{enhanc}
C.V. Johnson, A.W. Peet and J. Polchinski, \emph{Gauge theory and
the excision of repulson singularities}, Phys. Rev {\bf D61}
(2000) 086001, {\tt hep-th/9911161}; A. Buchel, A.W. Peet and J.
Polchinski, \emph{Gauge dual and noncommutative extension of an N
= 2 supergravity  solution}, Phys. Rev {\bf D63} (2001) 044009,
{\tt hep-th/0008076}; N. Evans, C.V. Johnson and M. Petrini,
\emph{The enhan\c{c}on and N = 2 gauge theory/gravity RG flows} JHEP
{\bf 10} (2000) 022, {\tt hep-th/0008081}; C. V. Johnson, R. C.
Myers, A. W. Peet and   S.F. Ross, \emph{The Enhan\c{c}on and the
Consistency of Excision}, {\tt hep-th/0105159}.
%
\bibitem{das} K. Dasgupta and S. Mukhi,
\emph{Brane constructions, fractional branes and anti-de Sitter
domain walls}, JHEP {\bf 07} (1999) 008, {\tt hep-th/9904131}.
%
\bibitem{GRANA} M. Grana and J. Polchinski,
{\em Supersymmetric Three-Form Flux Perturbations on $AdS_5$},
Phys. Rev. {\bf D63} (2001) 026001, {\tt hep-th/0009211}.
%
\bibitem{GUB} S. Gubser,
{\em Supersymmetry and F-theory realization of the deformed
conifold with  three-form flux}, {\tt hep-th/0010010}.
%
\bibitem{CVETIC} M. Cvetic, H. Lu and C.N. Pope,
{\em Brane resolution through transgression}, Nucl. Phys. {\bf
B600} (2001) 103, {\tt hep-th/0011023}.
%
\bibitem{CVETIC2}M. Cvetic, G. W. Gibbons, H. Lu and C. N. Pope,
{\em Ricci-flat metrics, harmonic forms and brane resolutions},
{\tt hep-th/0012011}.
%
\bibitem{ke}
A.~Kehagias, \emph{New type IIB vacua and their F-theory
interpretation}, Phys.\ Lett. {\bf B435} (1998) 337, {\tt
hep-th/9805131}.
%
\bibitem{GRANA2}
M. Grana and J. Polchinski,
{\em Gauge / gravity duals with holomorphic dilaton}, \hfill\break
{\tt hep-th/0106014}.
%
\bibitem{Schwarz:1983qr}
J.~H.~Schwarz, \emph{Covariant Field Equations Of Chiral N=2 D =
10 Supergravity}, Nucl.\ Phys. {\bf B226} (1983) 269.
%
\bibitem{igorleo} L.~Castellani and I.~Pesando,
\emph{The Complete superspace action of chiral D = 10, N=2
supergravity}, Int.\ J.\ Mod.\ Phys. {\bf A8} (1993) 1125.
%
\bibitem{castdauriafre} L. Castellani, R. D'Auria and P. Fr\'e, \emph{
Supergravity and Superstring theory: a geometric perspective},
World Scientific, Singapore (1990).
%
\bibitem{Becker} K. Becker and M. Becker,
\emph{M-Theory on Eight-Manifolds}, Nucl.Phys. {\bf B477} 155 (1996),
{\tt hep-th/9605053}.
%
\bibitem{mackay} J. Mc Kay in Proc. Symp. Pure Math., Am. Math. Soc. v{\bf
37} (1980) 183; P.B.Kronheimer, J. Differ. Geometry {\bf 29}
(1989) 665, J. Differ. Geometry {\bf 29} (1989) 685.
%
\bibitem{Eguchi:1978xp}
T.~Eguchi and A.~J.~Hanson, \emph{Asymptotically Flat Selfdual
Solutions To Euclidean Gravity}, Phys.\ Lett. {\bf B74} (1978)
249.
%
\bibitem{dennery} P. Dennery, A. Krzywicki, \emph{Mathematics for
Physicists}, Harper and Row 1967; \\E. L. Ince, \emph{Ordinary
Differential Equations}, Dover 1956.
%
\bibitem{popino1} H. L\"u, C.N. Pope, P.K. Townsend {\it Domain Walls
form Anti de Sitter space} Phys.Lett. {\bf B391} (1997) 39,
{\tt hep-th/9607164}.
%
\bibitem{parilez}
P.~Fre, \emph{ Gaugings and other supergravity tools of p-brane
physics}, {\tt hep-th/0102114}.
%
\bibitem{Boonstra:1999mp}
H.~J.~Boonstra, K.~Skenderis and P.~K.~Townsend, {\it The domain
wall/QFT correspondence} JHEP{\bf 01} (1999) 003, {\tt hep-th/9807137}.
%
\bibitem{fu}
F.~Fucito, J.~F.~Morales and A.~Tanzini,
\emph{D-instanton probes of non-conformal geometries},
{\tt hep-th/0106061}.
%
\bibitem{dario}
J.~P.~Gauntlett, N.~Kim, D.~Martelli and D.~Waldram,
\emph{Wrapped fivebranes and N=2 super Yang--Mills theory},
{\tt hep-th/0106117}.
%
\bibitem{zaffa}
F. Bigazzi, A. L. Cotrone, A. Zaffaroni,
\emph{N=2 Gauge Theories from Wrapped Five-branes}
{\tt hep-th/0106160}.
%
\end{thebibliography}
\end{document}